\documentstyle[twoside,fleqn,epsf,espcrc2]{article}
\begin{document}
\title{
\vspace*{-1cm}
Different Traces of Quantum Systems Having the Same Classical Limit}
\author{Prot Pako{\'n}ski\address{Instytut Fizyki
	im. Mariana Smoluchowskiego Uniwersytet Jagiello{\'n}ski,\\
	ul. Reymonta 4, 30--059 Krak{\'o}w, Poland\\
	e-mail: \emph{pakonski@if.uj.edu.pl}}}

\begin{abstract}
\vspace*{-0.3cm}
\hrule\vspace*{0.1cm}\noindent{\bf Abstract}\\[0.1cm]
Many quantum systems may have the same classical limit.
We argue that in the classical limit their traces do not
necessarily converge one to another. The trace formula allows
to express quantum traces by means of classical quantities
as sums over periodic orbits of the classical system. To
explain the lack of convergence of the traces we need
the quantum corrections to the classical actions of
periodic orbits. The four versions of the quantum baker map on
the sphere serve as an illustration of this problem.
\\[0.1cm] {\em PACS}: 05.45.Mt \\[0.05cm]
\hrule
\vspace*{-0.7cm}
\end{abstract}

\maketitle

\vspace*{-0.4cm}
\section{ Introduction } \label{Intro}
\vspace*{-0.3cm}
The analysis of a quantum system in the semiclassical regime
allows one to approximate quantum traces by the sum over periodic
orbits of the corresponding classical system. This relation
is provided by the famous Gutzwiller Trace Formula~\cite{Gu90}
which establishes a fruitful link between a quantum system and
its classical counterpart. It is well known that infinitely many
quantum systems may have the same classical limit. A natural
problem arises how different such quantum systems might be.
Let us concentrate on quantum traces which allow us to compute
the spectrum. The Gutzwiller-like formula approximates in the
semiclassical limit the traces of the evolution operator $U$ by
\begin{equation}
  \mbox{Tr}\ U^n = \sum_{\rm p.o.} A_o \cdot e^{i S_o / \hbar} \ .
    \label{TrForm}
\end{equation}
The sum is taken over all periodic orbits with periods equal
or less than $n$. The orbits contribute with amplitudes $A_o$
which depend on their stability while the phases are equal to
their classical actions (measured in $\hbar$ units). Possible Maslov
indices, not mentioned in the formula (\ref{TrForm}), do not
influence the following argument since they are related to the
classical system. An example of two quantum systems, with the same
classical limit, having the trace of the Floquet operator
constant with respect to $\hbar$ and equal to $0$ and $\sqrt{2}$
respectively is provided by the quantum baker map on the
sphere~\cite{POZ99}. To approximate these traces semiclassically
in the spirit of~(\ref{TrForm}) $\hbar$ corrections to the
classical actions, $S_o=S_{\rm class}+\hbar S_{\rm quant}$, are needed.
These corrections disappear in the classical limit $\hbar\rightarrow
0$, but they result in corrections of order unity to the traces.
The traces of such quantum systems can not be expressed using
only classical quantities and may not converge in the classical
limit. Many quantum models for which the semiclassical treatment
is known~\cite{T83,CT91,SV92,CTH92,GHKWZ95,BGHS96,KH98}
do not seem to suffer from this problem --- their traces do converge
to those predicted by the sum over periodic orbits without any
quantum corrections. An example where such $\hbar$ corrections are
needed in the semiclassical analysis is the original version of the
kicked top~\cite{KHE93}. On the other hand a small change of this
quantum model~\cite{BGHS96} is sufficient to eliminate this problem.

The Fourier transform with respect to $1/\hbar$ of the quantity
$\langle\Psi_0|U^n|\Psi_0\rangle$ for some state $|\Psi_0\rangle$
may be used to check the semiclassical predictions.
The modulus of this transform has peaks at the classical
actions of orbits of $n$ iterations of the map, if the state
$|\Psi_0\rangle$ overlaps with the state localized on the periodic
orbit. The phase of the transform will contain Maslov indices and
first order quantum corrections to the actions of periodic orbits.
Contributions stemming from periodic points with the same classical
action form single peak in the transform. In this work we propose
a way to analyze the classical action and its quantum correction
coming from a single periodic point of $n$ iterations of the map.
The paper is organized as follows. The decomposition of the Husimi
representation of the evolution operator into a sum over periodic
orbit is presented in section~\ref{Decomp}. The analysis of
actions of periodic orbits and their corrections found for the
quantum models of the baker map on the sphere is performed in
section~\ref{Baker}. Conclusions and consequences for the spectra
are discussed in section~\ref{Concl}.

\section{ Coherent State Decomposition of the Evolution Operator }
\vspace*{-0.3cm}
\label{Decomp}
Lets $\Omega$ be a classical phase space. Consider a classical
area-preserving map $\Theta: \Omega\rightarrow\Omega$ and a corresponding
quantum operator $U$ acting on a Hilbert space $\mathcal{H}$. Assume
we have a family of generalized coherent states $|\alpha\rangle\in
\mathcal{H}$ for any $\alpha\in\Omega$ fulfilling the minimal
uncertainty relation. Such a family of states is found for several
examples of classical phase spaces. The coherent states form
an overcomplete basis allowing the decomposition of the identity
operator $\int_\Omega|\alpha\rangle\langle\alpha|d\alpha=\mathbf{1}$.
We introduce the generalized Husimi~\cite{Hu40} representation
of an operator $\mathcal{A}$
\begin{equation}
  H_{\mathcal{A}}(\alpha) = \langle\alpha|\mathcal{A}|\alpha\rangle .
\end{equation}
If $\mathcal{A}$ is a projection operator onto state $|\Psi\rangle$,
it reduces to the standard Q-representation of a pure state $H_
{|\Psi\rangle\langle\Psi|}(\alpha)=|\langle\alpha|\Psi\rangle|^2$.
Representing the identity operator as a sum over coherent states we have
\begin{equation}
  \int_{\Omega}H_{\mathcal{A}}(\alpha)d\alpha=\mbox{Tr}\,{\mathcal{A}},
  \ \ \int_{\Omega}H_{|\Psi\rangle\langle\Psi|}(\alpha)d\alpha=1.
\end{equation}

The family of coherent states allows us to establish a link between
classical and quantum dynamics~\cite{Ki88,SZ94}. If for any $\rho>0$
\begin{equation}
  \lim_{\hbar\rightarrow 0} \ \inf_{\alpha\in\Omega} \ \int_
    {C(\Theta(\alpha),\rho)}|\langle\alpha'|U|\alpha\rangle|^2d\alpha'=1 \ ,
  \label{LimForm}
\end{equation}
where $C(\Theta(\alpha),\rho)$ denotes a circle centered at
$\Theta(\alpha)$ with the radius $\rho$, we say that the
quantization of $\Theta$ into $U$ is \emph{regular} with respect
to the family of coherent states $|\alpha\rangle$. The formula
(\ref{LimForm}) means that $|\langle\alpha'|U|\alpha\rangle|^2$
tends to $\delta\left(\alpha'-\Theta(\alpha)\right)$ in the
classical limit, so the quantum state $U|\alpha\rangle$ is then
concentrated in the vicinity of the classical image $\Theta(\alpha)$.

Let us consider the Husimi representation of the evolution operator
$H_{U^n}(\alpha)=\langle\alpha|U^n|\alpha\rangle$. The above argument
shows that in the semiclassical regime the state $U^n|\alpha\rangle$
is concentrated at $\Theta^n(\alpha)$ and is close to $0$ everywhere
else. Therefore $H_{U^n}(\alpha)$ is localized on the periodic points
of classical map ($\Theta^n(\alpha)=\alpha$). This useful feature was
observed in~\cite{Sa90} and served to demonstrate classical to quantum
correspondence. The author of~\cite{Sa90} investigated only the squared
modulus of the function $H_{U^n}(\alpha)$.

We know that $\mbox{Tr}\ U^n=\int_\Omega\langle\alpha|U^n|\alpha\rangle
d\alpha$ and the integrated function is localized on the periodic
points in the classical limit ($\hbar\rightarrow 0$). This fact
allows us to decompose the quantum traces into sums over periodic
orbits for any quantum maps, under the standard assumption that
all periodic points are isolated, but without employing the saddle
point approximation. The modulus of the integrated overlap
$\langle\alpha|U^n|\alpha\rangle$ near the periodic point
corresponds to the amplitude $A_o$ in equation~(\ref{TrForm}) and
depends on the stability of this periodic orbit. The phase of the
function $H_{U^n}(\alpha)$ seems to have a stationary point at the
periodic orbit as observed in~\cite{KHE93} for the quantum kicked
top and reported later on in this work for the baker map on the sphere.

\section{ Various Quantizations of Baker Map on the Sphere } \label{Baker}
\vspace*{-0.3cm}
Quantizing the baker map on the sphere~\cite{POZ99} we
define four different quantum maps $B_{k}$ ($k=0\ldots 3$),
corresponding to the same classical system. In spite of
this fact, they have different traces and spectra. By a
quantum map on the sphere we understand a $N=2j+1$ dimensional
unitary operator acting in the Hilbert space of angular momentum,
where $j(j+1)$ is the eigenvalue of $\hat{J}^2$ operator which is
conserved by the evolution. The four unitary operators $B_k$ may
be expressed in the eigenbasis $|j,m\rangle$ of $\hat{J}_z$ operator
\begin{equation}
  B_{(ab)} = R^{-1}\left[\begin{array}{cc}R'^{(a)} & 0 \\
    0 & R''^{(b)} \end{array}\right] \ , \label{SphBak1}
\end{equation}
where index $(ab)$ with $a,b=0,1$ corresponds to binary representation
of $k$ ($B_{(00)}=B_0$, etc.), $R$ is the Wigner rotation matrix
$R_{m',m} =\langle j,m|e^{-i\frac{\pi}{2}\hat{J}_y}|j,m'\rangle$
and the matrices $R'$ and $R''$ are constructed by choosing odd
or even columns of the matrix $R$ in the following way
\[
  {R'}^{(a)}_{m,l} := \sqrt{2}R_{m,2l+j+a}, \ \ \ m,l=-j\ldots -\frac{1}{2};
\]
\begin{equation}
  {R''}^{(b)}_{m,l} := \sqrt{2}R_{m,2l-j-1+b},
  \ \ \ m,l=\frac{1}{2}\ldots j. \label{SphBak2}
\end{equation}
This construction requires the dimension $N$ to be an even
integer so the quantum number $j$ is half integer. The volume
of the classical phase space is equal to $4\pi$ (the unit sphere),
so $N=\frac{4\pi}{2\pi\hbar}$, which relates $\hbar$ to the
size of the Hilbert space ($\hbar=2/N$) and the classical
limit corresponds to $N\rightarrow\infty$. Now the coherent
state $|\alpha\rangle$ will refer to a $SU(2)$ vector coherent
state $|\theta,\phi\rangle$~\cite{Pe86,VS95}.

\begin{figure}[htb]
  \vspace{-0.9cm}
  \epsfxsize=0.46\textwidth
  \epsfbox{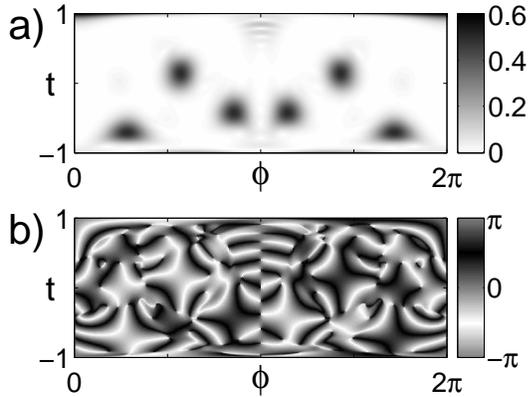}
  \vspace{-1.1cm}
  \caption{ The Husimi representation of three step propagator
	    for the baker map on the sphere is plotted using
	    gray scale (the coordinates are azimuth angle $\phi$
	    and $t=\cos\theta$), a) represents the absolute value
	    and b) the phase of the function $\langle\alpha|
	    (B_1)^3|\alpha\rangle$. Note the stationary phase
	    at the points where the modulus of the function
	    is maximal. \label{FigHFun} }
  \vspace{-0.8cm}
\end{figure}

In figure~\ref{FigHFun} we plot the Husimi representation of the
operator $(B_1)^3$ for $N=100$, the coordinates are the azimuth angle
$\phi$ and $t=\cos\theta$. The complex function $\langle\alpha|
(B_1)^3|\alpha\rangle$ is well localized near the periodic points
of three iterations of the classical baker map on the sphere. The
analogous plots for other versions of the model look almost the
same. We want to investigate the phase of contribution coming
from single periodic point. The numerical integration performed
for several dimensions $N$ shows that the phases of such
contributions are very well approximated by
$\arg(\langle\alpha_o|(B_k)^n|\alpha_o\rangle)$, where $|\alpha_o
\rangle$ is the coherent state centered at the periodic point.
This is so due to the stationarity of the phase of the function
$\langle\alpha|(B_k)^n|\alpha\rangle$ at the periodic point, as
pointed out before.

\begin{figure}[htb]
  \vspace{-0.9cm}
  \epsfxsize=0.46\textwidth
  \epsfbox{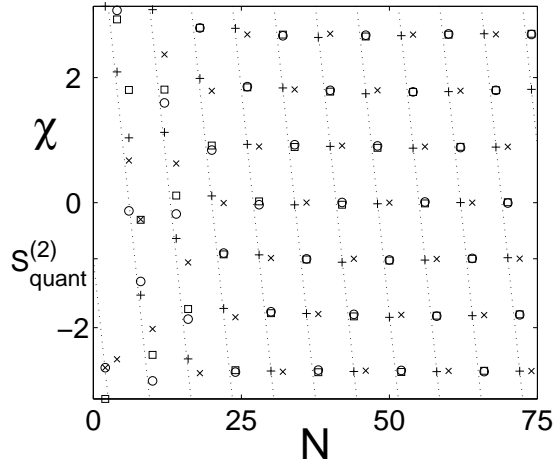}
  \vspace{-1.1cm}
  \caption{ The phase $\chi_k=\arg(\langle\alpha_o|(B_k)^3|\alpha_o
	    \rangle)$ for the baker map on the sphere is plotted
	    as a function of $N=2/\hbar$ the dimension of Hilbert
	    space. The four versions of the model $k=0\ldots 3$
	    are marked by $\circ, \times, +, \Box$ respectively.
	    Note the linear behavior for larger value of $N$, the
	    dotted line ($\chi=-\frac{2}{7}\pi N-\frac{2}{7}\pi$)
	    represents the asymptotic behavior of one of the models
	    --- $B_2$ ($+$). The quantum correction $S_{\rm quant}^{(2)}$
	    corresponds to the point, where this line crosses the $N=0$
	    axes. The coherent state $|\alpha_o\rangle$ is chosen at
	    the periodic point $\theta=\arccos\left(-\frac{5}{7}\right)$,
	    $\phi=\frac{2}{7}\pi$. \label{FigHAct} }
  \vspace{-0.8cm}
\end{figure}

We have calculated the phase of the Husimi representation of $(B_k)^3$
for four versions of the model defined by equations~(\ref{SphBak1})
and~(\ref{SphBak2}) at the periodic point $\theta=\arccos\left(-
\frac{5}{7}\right)$, $\phi=\frac{2}{7}\pi$. The results are plotted
in figure~\ref{FigHAct} where the four different symbols denote the four
different quantum maps. The phase $\chi$ is plotted in the
interval $(-\pi,\pi]$. We observe that, starting from some
dimension, the phase depends linearly on $N$ modulo $2\pi$.
The points for four versions of the model, marked with different
symbols, form parallel lines and they do not converge one to
another. The phase $\chi_k=\arg(\langle\alpha_o|(B_k)^3|
\alpha_o\rangle)$ can be then well parametrized by
$S_{\rm class}\frac{N}{2}+S_{\rm quant}^{(k)}$. The slope of
these lines is the same and corresponds to the classical action
of the orbit $S_{\rm class}$ equal to $-\frac {2}{7}2\pi$ (only the
fractional part of $2\pi$ is important here). The quantum
corrections $S_{\rm quant}$ to the actions of periodic orbits may be
found by an extrapolation of the linear behavior to $N=0$, as it
is demonstrated in figure~\ref{FigHAct} for $k=2$. The asymptotic
dotted line crosses $N=0$ axes at $\chi=S_{\rm quant}^{(2)}=-\frac{2}
{7}\pi$. The quantum corrections for other quantizations are
$S_{\rm quant}^{(0)}=S_{\rm quant}^{(3)}=0$ and
$S_{\rm quant}^{(1)}=\frac{2}{7}\pi$.

The investigation performed for other periodic points shows
that the quantum corrections $S_{\rm quant}$ are the same for
periodic points belonging to the same periodic orbit and
are equal to $p\times S_{\rm quant}$(primary orbit) if the
periodic orbit is $p$ repetitions of a primary orbit,
therefore with this respect they behave like classical
actions.

\vspace*{-0.3cm}
\section{ Conclusions } \label{Concl}
\vspace*{-0.3cm}
In this paper we have focussed on the semiclassical behavior of
quantum systems having the same classical limit. The corrections to
the quantum traces, not disappearing in the classical limit, arise
from $\hbar$ corrections to the actions of periodic orbits which
do disappear in the classical limit. It means that the traces of a
quantum system having such corrections may not be approximated
by the trace formula constructed on purely classical properties
of the system. For any quantum system we propose a way to find
the contributions to the traces coming from single periodic
points of $n$ iterations of the classical map by means of
autocorrelation function $\langle\alpha|U^n|\alpha\rangle$.
This function is localized at the periodic orbits of classical
system ($U$ stands for evolution operator and $|\alpha\rangle$
denotes the coherent state -- the wave packet). The four
versions of the quantum baker map on the sphere serve as an
example of such behavior. We investigate the phase $\chi=
\arg(\langle\alpha_o|B^n|\alpha_o\rangle)$ for the coherent state
$|\alpha_o\rangle$ pointing at the periodic points of the classical
baker map on the sphere. We have observed the linear behavior of $\chi$
as a function of $1/\hbar$ in semiclassical regime.

We believe that quantum systems having the same classical limit
may have different traces not converging one to another in the
classical limit. These traces can be explained by means of quantum
corrections (first order in $\hbar$) to the actions of periodic
orbits in the semiclassical approximation. The spectra of the
Floquet operators of such systems contain $N\sim1/\hbar$ eigenvalues
placed on the unit circle. They may be expressed in terms of the
traces. The difference of quantum traces of order of $1$ requires
difference between eigenvalues of different quantum systems at
least of order of $1/N$, i.e. the same as the mean level spacing.
The arguments presented in this paper should be valid even in the
case of infinite Hilbert space (e.g. for continuous dynamics).

I acknowledge fruitful discussions with \linebreak F.~Haake and
K.~{\.Z}yczkowski. It is a pleasure to thank the organizers of
the workshop \emph{Dynamics of Complex Systems} for the
invitation to Dresden.


\begin{thebibliography}{00}
  \vspace*{-0.3cm}
  \bibitem{Gu90}
    M.~C.~Gutzwiller, Chaos in Classical and Quantum Mechanics,
    Springer, Berlin, 1990.
  \bibitem{POZ99}
    P.~Pako{\'n}ski, A.~Ostruszka, K.~{\.Z}yczkowski,
    Nonlinearity 12 (1999) 269.
  \bibitem{T83}
    M.~Tabor, Physica D 6 (1983) 195.
  \bibitem{CT91}
    P.~W.~O'Connor, S.~Tomsovic, Ann. Phys. NY 207 (1991) 218.
  \bibitem{SV92}
    M.~Saraceno, A.~Voros, Chaos 2 (1992) 99, Physica D 79 (1994) 206.
  \bibitem{CTH92}
    P.~W.~O'Connor, S.~Tomsovic, E.~J.~Heller, J. Stat. Phys. 68 (1992) 131,
    Physica D 55 (1992) 340.
  \bibitem{GHKWZ95}
    P.~Gerwinski, F.~Haake, M.~Ku{\'s}, H.~Wiedemann, K.~{\.Zyczkowski},
    Phys. Rev. Lett. 74 (1995) 1562.
  \bibitem{BGHS96}
    P.~A.~Braun, P.~Gerwinski, F.~Haake, H.~Schomerus,
    Z. Phys. B 100 (1996) 115.
  \bibitem{KH98}
    L.~Kaplan, E.~J.~Heller, Phys. Rev. Lett. 76 (1998) 1453.
  \bibitem{KHE93}
    M.~Ku\'s, F.~Haake, B.~Eckhardt,
    Z. Phys. B 92 (1993) 221.
  \bibitem{Hu40}
    K.~Husimi, Proc. Phys. Math. Soc. Jpn. 22 (1940) 264.
  \bibitem{Ki88}
    Y.~Kifer, Random Perturbations of Dynamical Systems, Birkh{\"a}user,
    Boston, 1988
  \bibitem{SZ94}
    W.~S{\l}omczy{\'n}ski, K.~{\.Z}yczkowski, J. Math. Phys. 35 (1994) 5674.
  \bibitem{Sa90}
    M.~Saraceno, Ann. Phys. 199 (1990) 37.
  \bibitem{Pe86}
    A.~Perelomov, Generalized Coherent States and their Applications, Springer,
    Berlin 1986.
  \bibitem{VS95}
    V.~R.~Vieira, P.~D.~Sacramento, Ann. Phys. NY 242 (1995) 188.
\end{thebibliography}
\end{document}